# A pH-based bio-rheostat: a proof-of-concept

E. Alfinito,[1,a,b)] R. Cataldo[1,b] and L. Reggiani[1,b)]

[1]*Department of Mathematics and Physics, "Ennio De Giorgi" University of Salento, Lecce, I-73100, Italy*

New science and new technology need new materials and new concepts. In this respect, biological matter can play a primary role because it is a material with interesting and innovative features that have found several applications in technology, from highly sensitive sensors for medical treatments to devices for energy harvesting. Furthermore, most of its phenomenology remains unclear thus giving new hints for speculative investigations. In this letter, we explore the possibility to use a well-known photosensitive protein, the Reaction Center of *Rhodobacter Sphaeroides*, to build up an electrical pH sensor, i.e., a device able to change its resistance depending on the pH of the solution in which it crystalizes. By using a microscopic model successfully tested on analogue proteins, we investigate the electrical response of the Reaction Center single protein under different conditions of applied bias, showing the feasibility of the bio-rheostat hypothesis. As a matter of facts, the calculated resistance of this protein grows of about 100% when going from a pH = 10 to a pH = 6.5. Moreover, calculations of the current voltage characteristics well agree with available experiments performed with a current atomic force microscopy under neutral conditions. All findings are in qualitative agreement with the known role of pH in biochemical activities of Reaction Center and similar proteins, therefore supporting a proof-of-concept for the development of new electron devices based on biomaterials.

[a)] Author to whom correspondence should be addressed. Electronic mail: eleonora.alfinito@unisalento.it.

[b)] E. Alfinito designed research. R. Cataldo performed simulations, E. Alfinito and L. Reggiani performed research; all authors wrote paper.

Advances in electronics mainly demand for the development of devices eco-sustainable, able to operate under low-consumption and poor-waste conditions thus leading to fast moving consumer goods. In this perspective, increasing interest is recently addressed to the use of biomaterials. Few prominent examples can be found in the development of biobatteries [Chomicz et al. (2021)] or wearable biosensors [Hou et al. (2021)]. Among biomaterials, a primary role is played by sensing proteins, i.e., proteins able to convert the capture of an external agent (light or molecules) in a cascade of biochemical events leading to a significant change of their physical properties. Sensing proteins are characterized by a quite complex 3D structure that allows them to perform their natural sensing actions. These actions have been, at least partially, implemented in electronic devices and studied in terms of their related electrical properties [Alfinito et al. (2013), Alfinito & Reggiani (2015a)]. From one side, it has been observed that their integration in standard electronics could promote the efficiency and enlarge the field of applications of several devices [Li et al. (2018)]; on the other side, sensing proteins can be used like electronic components, independently from their specific natural function. This is the case of photosensitive proteins (PPs), i.e., proteins able to convert radiation in the visible range or in the near red/blue spectrum into chemical and electrical energy. The use of PPs is continuously implemented in biomedical applications, the best-known example being optogenetics [Deisserroth (2011)], as well as in devices for energy production [Sun et al. (2020)]. The PPs that received a relevant attention are some type-1 opsins, in particular bacteriorhodopsin (*bR*) and proteorhodopsin (*pR*), found in primeval organisms like archea and bacteria [Béja et al. (2000)]. Both these proteins are quite small (about 25 kDa in atomic mass) and coupled to a retinal molecule. After illumination, the coupled retinal molecule undergoes a conformational change, modifying the structure of the whole protein, and allowing the transfer of a proton across the cell membrane. More recently, increasing attention has been devoted to another PP that is of paramount interest in photosynthesis, the Reaction Center (*RC*), found in several sulphur and not sulphur bacteria like the *Rhodobacter Spaeroides*. This protein is quite large (about 91 kDa) and is coupled to different elements: pigments, light antennas and two quinones ($Q_A$ and $Q_B$) that assist the charge transfer through the whole protein. *RC* structure is significantly more complex than that of *pR* and *bR*. The same occurs for its biochemical activity that includes a light activated mechanism of charge separation and electron transport [Allen et al. (1987); Xu et al. (2001); Tamura et al. (2021)]. The mechanism of light harvesting that *RC* implements has inspired some kinds of photocells [O'Regan & Grätzel (1991)] and has stimulated interest for the development of the 4th-generation photovoltaics. Indeed, *RC* arranges its 3D structure following the environmental pH value. Consequently, this protein is a promising candidate to be implemented in devices like pH-sensors or bio-rheostats. The implications and the mechanisms responsible of this implementation have been also observed in other proteins, and the associated modifications could be of sufficiently relevance to eventually drive the protein toward denaturation. Despite the relevance of these effects, these phenomena are not completely clarified [Srivastava et al. (2007)] and should deserve further investigations.

Several crystallographic freeze-trapping experiments carried out on *RC* suggested a significant conformational change within the secondary quinone ($Q_B$) binding site in response to an electron transfer. $Q_B$ binds in a "distal" binding site in dark and moves approximately 4.5 Å to a "proximal" binding site upon illumination [Baxter et al. (2004)]. The existence of two different binding sites is confirmed in [Koepke (2007)], that performed a quite complete study to find out a relation between the percentage of $Q_B$s in the proximal configuration and the value of pH, also producing a valuable dataset deposited on the Protein Data Bank (PDB) [Berman et al. (2000)].

The aim of this letter is to investigate the sensing properties of the *RC* making use of the structures listed in Table I, describing the same protein crystallized in different values of pH, both in dark and light. To this purpose, we analyse the expected electrical properties of the *RC* protein, with the objective to quantitatively estimate the effects of a pH variation on the electrical response. Investigation is performed in the framework of Proteotronics [Alfinito et al. (2015b)], a recent branch of electronics devoted to investigate the structural and functional properties of sensing proteins by using electronic methods. The model focuses on the role of the primary structure and its conformational modifications in the detectable electrical properties. To this purposes, the model only accounts for the amino acid tertiary structures and their specific electrical properties.

In brief, the protein is represented by a set of nodes, one for each considered amino acid, and a set of links each connecting a couple of nodes when their distance, $l_{i,j}$, is below an appropriate cut-off value ($D$).[Alfinito et al. ,2015] In this way, the complete graph representing the macromolecule is described by a Boolean matrix, $B_{ij}$, 1/0 corresponding to the presence or any of a link between nodes *i* and *j*.

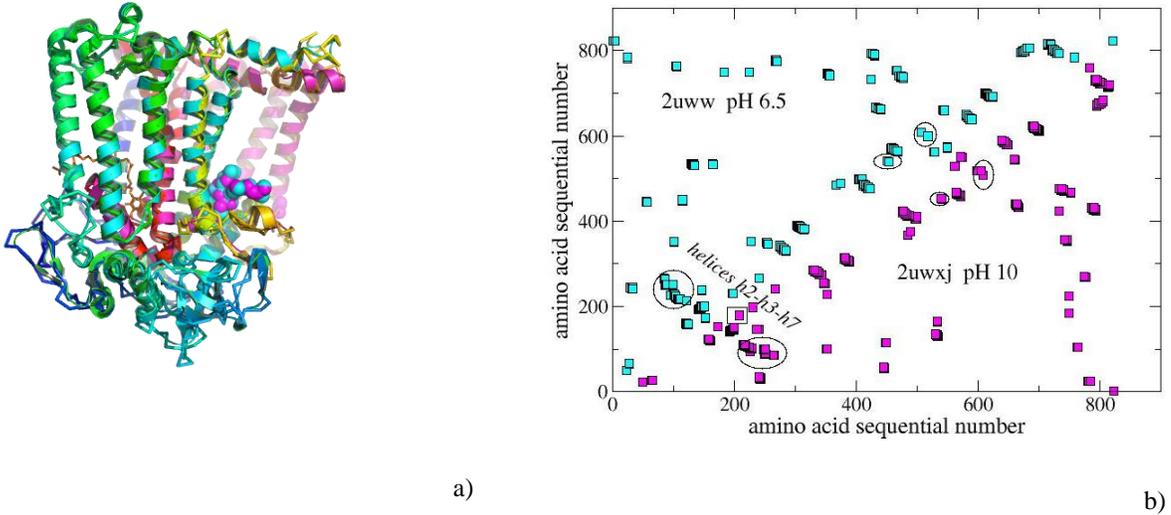

FIG. 1. Different protein representations. a) the structures of *RC* apoprotein and the quinones are superimposed; $Q_A$ is in orange sticks, $Q_B$ is in balls, cyan for 2uww and magenta for 2uxj [PyMol]; b) the corresponding contact maps, for 2uww (on the left), and for 2uwj (on the right), obtained for $D$=6 Å. Major differences are highlighted by closed circles.

Figures 1 reports two different visualization of the protein in dark at pH = 6.5 and pH = 10: In Fig. (1a) the (3D) cartoon of RC is drawn; in Fig. (1b), the (2D) protein contact maps are shown. A contact map represents the off-diagonal parts of the matrix $B_{ij}$ and accounts for the connected pairs of nodes. The structures taken at different pH values have tiny differences that, in the contact maps, are encapsulated in small circle /boxes. In the present case, we notice displacements of helices 2,3,7 of chain H as well as in the terminal part of chain L.

For the calculations of the electrical properties, each link is interpreted as a channel for charge transfer whose resistance $R_{i,j}$ is taken to be in general function of the geometry and of the link resistivity ρ as [Alfinito et al. (2011); Alfinito et al. (2015b); Alfinito et al. (2016)]:

$$R_{i,j} = \rho \frac{4 \, l_{i,j}}{\pi(D^2 - l_{i,j}^2)} \qquad (1)$$

The resistance network resulting from the set of links and nodes preserves the main features of the protein that are relevant for calculations, i.e., the amino acid configuration and the electrical properties of each link. By using Kirchhoff laws, the charge transfer inside the network is described by a set of linear equations that are solved by a standard numerical procedure. The solution so obtained, provides the electrical response of the network in terms of its local currents and voltage drops, its global resistance and/or its static current-voltage characteristics. We remark that different 3D structures produce different electrical outputs which can be compared with what found in experiments performed in light and in dark in *bR* or *pR* [Jin et al. (2008), Melikyan et al. (2011)], where a marked difference in current was observed because of the light induced conformational change [Alfinito et al. (2013), Alfinito & Reggiani (2014)].

In the present case of *RC*, in dark, a pH change produces a re-alignment of QBs which propagates to the whole structure, also changing the global protein resistance.

The expected resistance change is evaluated by using different resistor networks for *RC* in dark pertaining to different pH conditions. Each of these networks comes from the appropriate structures, listed in Table 1, and their performances are described in Fig. 2 and Fig. 3, by using as a benchmark the structure in dark at pH 6.5 whose PDB entry is *2uww*.

Figure 2 reports the relative resistance (the resistance normalized to that of the benchmark) of the structures taken in dark as a function of the cut-off distance between neighborhood nodes, $D$. By increasing the pH value, the protein structure changes and the corresponding resistance is found to become smaller than that of the benchmark, thus suggesting the occurrence of a reduction of the protein **size** at increasing the pH value of the solution. At $D = 10.5$ Å, resistance monotonically reduces for pH values in the range from 8 to 10 up to about halving the value at pH = 6.5 (see Table I). By further increasing $D$, more and more links should be considered in the network and the corresponding network resistance tends to become constant when practically each node is linked to all the other nodes.

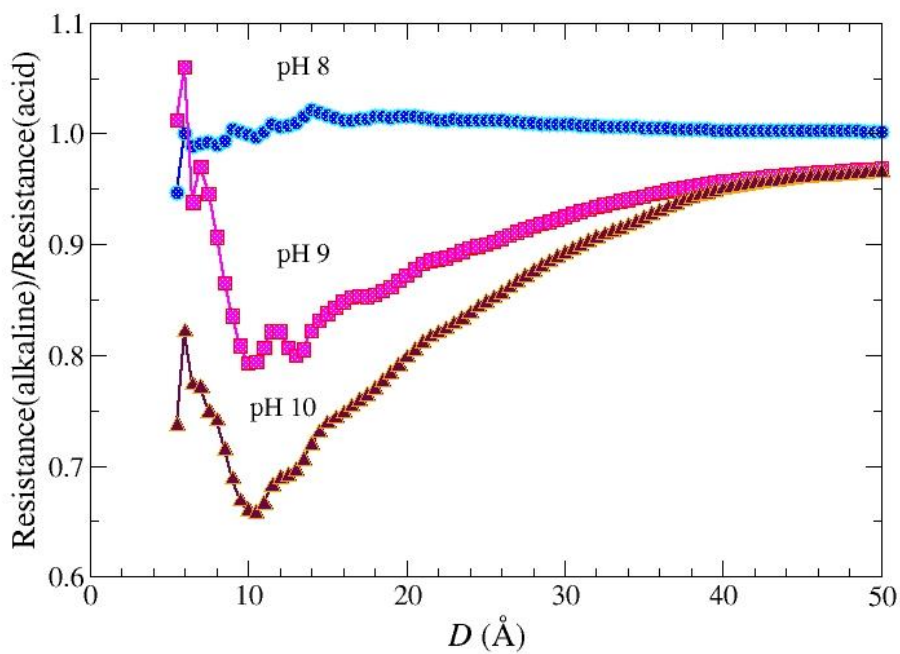

FIG. 2. Relative resistances of the single protein structures in dark. The resistance of each structure is calculated as function of the cut-off distance and compared with that of 2uww, the in-dark structure taken as a benchmark, obtained in weak acid conditions of pH=6.5. Calculations are carried out by taking the same value for the resistivity. Symbols refer to the calculated values; dotted lines are guides for the eye.

It is reasonable to assume that this behaviour is correlated with a decrease of the $Q_B$ population in the proximal configuration [Koepke et al. (2007)] (see also Table I), thus being the main responsible of the protein conformational change. In [Koepke et al. (2007)], the secondary quinones appear uniformly distributed between the proximal and distal configurations, under weak acid condition; and, when passing to the alkaline region, an abrupt reorientation is observed, which relaxes at further increasing pH values. The origin of this reorientation is not clear: it could signal a change in the protein functioning like, for example, an inversion of vectoriality in the proton pumping when going from the weak-acid (or neutral) conditions to the alkaline ones [Altamura et al. (2018)]. Indeed, a similar behavior has been observed also in pR [Friedrich et al. (2002)].

Finally, we calculate the expected I-V characteristics of the single protein in dark, respectively for pH = 6.5 and pH = 10. To reproduce the superohmic behavior exhibited by experiments at increasing values of an applied positive-voltage (Mikayama et al., 2008), the microscopic model makes use of a sequential tunneling mechanism of charge transfer between different nodes that are assumed to be separated by an energy barrier $\Phi$. Specifically, the current response is simulated by using a Monte Carlo procedure to allow the charge transfer channels to reduce their initial resistivity from a $\rho_{MAX}$ value to a $\rho_{min}$ value at increasing values of the potential drop between nodes [Alfinito et al. (2011), Alfinito & Reggiani (2015c)]. This is on the wake of the well-known Simmons model for the charge injection in an electronic junction (Simmons, 1963). This model describes two different tunnelling regimes: at low bias it envisages a direct tunnelling mechanism, at high bias overtaken by an injection tunnelling mechanism. At low bias, the condition e $V_{i,j} < \Phi$ holds for most of the channels that take the same $\rho_{MAX}$ value and the global response of the protein will be similar to that of an insulator. At increasing bias, when $eV_{i,j} > \Phi$ an abrupt jump of resistivity to a minimal value, $\rho_{min}$, occurs for the given channel and the global resistance of the protein will decrease accordingly. At further increasing of the bias, most of the channels will take the $\rho_{min}$ value and the global response of the protein will be similar to that of a conductor. The tunnelling transition probabilities including direct and injection mechanisms write:

$$P_{i,j}^D = \exp{-\beta \left(\Phi - \frac{eV_{i,j}}{2}\right)}, \qquad if \ eV_{i,j} < \Phi \qquad (2a)$$

$$P_{i,j}^I = \exp{-\beta \left(\frac{\Phi^{3/2}}{\sqrt{2}eV_{i,j}}\right)}, \qquad if \ eV_{i,j} \geq \Phi \qquad (2b)$$

where $\beta = \frac{l_{i,j}\sqrt{8m}}{\hbar}$, with h the Planck constant, accounts for the effective electron mass (m) and of the distance ($l_{i,j}$) between the considered nodes.

Here, by using a barrier height of 0.219 eV and resistivity ρ in the range [$10^{11}$-$10^4$] ΩÅ we simulate the current response as given in (Mikayama et al., 2008). Figure 3 reports the results of calculations. Specifically, Fig. 3a compares theoretical values calculated for a pH=6.5 structure with experiments performed in dark and in a similar pH = 7.5 condition when the protein is sandwiched between two different kinds of modified substrates. Each substrate specifically affects the intensity of the measured current. In our modelling this difference in the substrates is accounted for by using two different values of $\rho_{MAX}$, respectively $\rho_{MAX}$ =5.4e10, and 1.2e11 ΩÅ, although preserving the value of the barrier height, and of $\rho_{min}$ = 2.7 $10^4$ ΩÅ. Figure 3b reports the expected current response for the same protein in dark and under near neutral and strong alkaline conditions. The significant increase of the current for the alkaline conditions agrees qualitatively with the decrease of resistance evidenced in the literature (Ohno et al, 2009), even if direct I-V experiments are not available at present. In both the cases here considered, theoretical calculations reported in Fig. 3 well compare with available experiments thus supporting the physical reliability of the microscopic model.

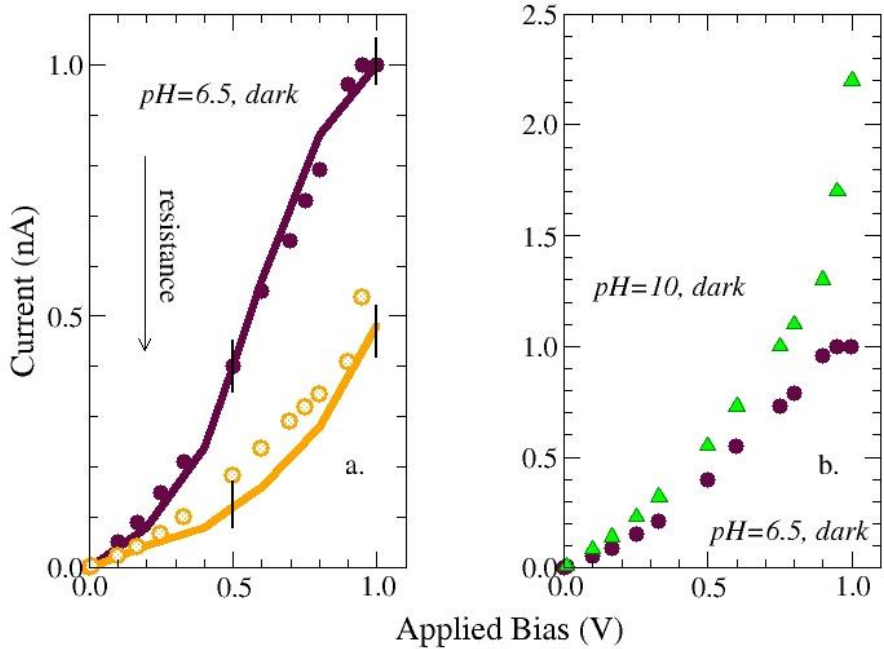

FIG. 3. Calculated I-V for the dark-adapted structures of *RC* . a. Simulations refer to resistivity values, $\rho_{MAX} = 5.4 \; 10^{10} \; \Omega\text{Å}$ (full circles), $\rho_{MAX} = 1.2 \; 10^{11} \; \Omega\text{Å}$ (open circles) and $\rho_{min} = 2.7 \; 10^{4} \; \Omega\text{Å}$. Comparison is made with the data obtained in (Mikayama et al., 2008) using Au/2MP/RC/2MP/Au and Au/2MP/RC/Au configuration (continuous lines). b) Simulations refer to resistivity values, $\rho_{MAX} = 5.4 \; 10^{10} \; \Omega\text{Å}$ and $\rho_{min} = 2.7 \; 10^{4} \; \Omega\text{Å}$ with two different pH value (pH 6.5, circles) and (pH 10, triangles). Data refer to mean values each calculated on a single Monte Carlo realization of $2 \times 10^{4}$ iterations.

TABLE I. List of the structures used in the present analysis.

| PDB[a] | %[b] | rr[c] | pH[d] |
|--------|------|-------|-------|
| 2uww   | 55   | 1     | 6.5   |
| 2j8c   | 65   | 0.99  | 8     |
| 2ux3*  | 55   | 0.79  | 9     |
| 2uxj   | 35   | 0.66  | 10    |

[a] *PDB* is the PDB entry for the dark-adapted structures.
[b] *%* is the percentage of the secondary quinones in the proximal positions [Koepke et al. (2007)].
[c] *rr* is the resistance of the structure in the dark (at different pH values) vs the resistance of 2uww (calculated at *D*=10.5 Å).
[d] *pH* is the pH value [Koepke et al.(2007)].
* for pH 9 the structures were taken in different experiments.

In conclusion, by using the tertiary structure of the *RC* protein, its electrical characteristics are investigated in the presence of a pH of the crystalized solution ranging from weak acid up to strong alkaline values. Theoretical calculations are carried out in the framework of the Proteotronics, a structured approach to the new field of protein-based electronics, and parallels previous investigations performed for similar proteins belonging to the opsin family. Numerical results show that the *RC* protein is extremely sensitive to the pH value of the solution in which it is crystallized. Structural differences, that are not very evident to the naked eye [Srivastava et al. (2007)], are emphasized by the change in the single protein resistance. Looking at the relative resistance reported in Fig. 2, we could conclude that the protein undergoes a shrinkage when the pH changes from weak acid (pH = 6.5) to alkaline conditions, and by increasing the pH value this change appears even more evident. These results are supported by previous studies, that propose the resistance measurements as a privileged tool for the investigation of the protein structure. Furthermore, as a new result, we found that the sensitivity of *RC* strictly depends on pH, thus this protein could be used to implement a bio-rheostat, i.e. a device able to translate a change in pH into a change in resistance. *Vice versa, RC* could be used for sensor applications in testing the value of pH too. Consequently, present results provide a proof of concept, that could be easily generalized and useful for other proteins.

## ACKNOWLEDGMENTS

Profs M.R. Guascito and L. Giotta (Department of Biological and Environmental Sciences and Technologies, University of Salento) and dr. Francesco Milano (CNR-ISPA Istituto di Scienze delle Produzioni Alimentari, Consiglio Nazionale delle Ricerche) are gratefully acknowledged for useful discussions.

## AVAILABILITY OF DATA

The data that support the findings of this study are available from the corresponding author upon reasonable request.